\documentclass[10pt,conference,letterpaper]{IEEEtran}

\usepackage[pdftex]{graphicx}
\usepackage{amssymb}
\usepackage[vlined,linesnumbered,boxed]{algorithm2e}
\usepackage{times}
\usepackage{comment}
\usepackage{subfigure}

\begin{document}

\title{Reliable Provisioning of Spot Instances for Compute-intensive Applications}

\author{\IEEEauthorblockN{William Voorsluys and Rajkumar Buyya}
\IEEEauthorblockA{Cloud Computing and Distributed Systems (CLOUDS) Laboratory\\
Department of Computer Science and Software Engineering\\The University of Melbourne, Australia\\
Email:\{williamv,rbuyya\}@csse.unimelb.edu.au}
}

\maketitle

\begin{abstract}
Cloud computing providers are now offering their unused resources for leasing in the spot market, which has been considered the first step towards a full-fledged market economy for computational resources. Spot instances are virtual machines (VMs) available at lower prices than their standard on-demand counterparts. These VMs will run for as long as the current price is lower than the maximum bid price users are willing to pay per hour. Spot instances have been increasingly used for executing compute-intensive applications. In spite of an apparent economical advantage, due to an intermittent nature of biddable resources, application execution times may be prolonged or they may not finish at all. This paper proposes a resource allocation strategy that addresses the problem of running compute-intensive jobs on a pool of intermittent virtual machines, while also aiming to run applications in a fast and economical way. To mitigate potential unavailability periods, a multifaceted fault-aware resource provisioning policy is proposed. Our solution employs price and runtime estimation mechanisms, as well as three fault tolerance techniques, namely checkpointing, task duplication and migration. We evaluate our strategies using trace-driven simulations, which take as input real price variation traces, as well as an application trace from the Parallel Workload Archive. Our results demonstrate the effectiveness of executing applications on spot instances, respecting QoS constraints, despite occasional failures.
\end{abstract}

\section{Introduction}
\label{intro}

Variable pricing virtual machines (also know as ``spot instances''\footnote{The terms ``spot instance'', ``instance'', ``virtual machine'', ``VM'', and ``resource'' signify the same concept and are used interchangeably in this work.}) are increasingly being employed as a means of accomplishing various computational tasks, including high performance parallel processing tasks, which are common in several areas of science, such as climate modeling, drug design, and protein analysis, as well in data analytics scenarios, such as execution of MapReduce tasks~\cite{chohan2010see}.  Significant cost savings and the possibility of easily leasing extra resources when needed, are major considerations when choosing virtual clusters, dynamically assembled out of cloud computing resources, over a local HPC cluster~\cite{voorsluys2011spot}.

The cloud computing spot market, since introduced by Amazon Web Services~\cite{aws, varia2011best}, has been considered as the first step for a full-fledged market economy for computational resources~\cite{zhang2011}. In this market, users submit a resource leasing request that specifies a maximum price (bid) they are willing to pay per hour for a predefined instance type. Instances associated to that request will run for as long as the current spot price is lower than the specified bid. Prices vary frequently, based on supply and demand. Price are distinct and vary independently for each available datacenter (``availability zone'' in Amazon terminology), spot instance type, and operating system choice. Not all type/OS combinations are available in all datacenters. In other words, there are multiple spot markets from where to choose suitable computational resources, making the provisioning problem significantly challenging.

When an out-of-bid situation occurs, i.e. the current spot price for that instance type goes above the user's maximum bid, instances are terminated by the provider without prior notice. Therefore, in spite of an apparent economical advantage, an intermittent nature is inherent to biddable resources, which may cause VM unavailability.

Despite the possibility of failures due to out-of-bid situations, as we have discussed in our previous work~\cite{voorsluys2011spot}, it is advantageous to utilize spot instances to run compute-intensive applications at a fraction of the price that would normally cost when using standard fixed-priced VMs. Specifically, we have demonstrated the effect of different runtime estimation methods on the decision-making process of a dynamic job allocation policy. Our policy was responsible for requesting and terminating spot instances on-the-fly as needed by a stream of computational jobs, as well as choosing the best instance type for each job based on the estimated job execution time on each available type.

We had previously assumed that users would bid high enough so that the chance of spot instance failures due to out-of-bid situations would be negligible. In reality, even though users only pay the current spot price at the beginning of each hour, regardless of the specified bid, there are incentives for bidding lower. Andrzejak et al, who evaluated checkpointing techniques for spot instance fault tolerance, observed that by bidding low, significant cost savings can be achieved, but execution times increase significantly. Similarly, by increasing the budget slightly, execution times can be reduced by a large factor~\cite{Andrzejak2010}.

\subsection{Bidding strategies and the need for fault tolerance}

We now elaborate on the potential risks and rewards of provisioning a resource pool composed exclusively of spot instances in scenarios where QoS constraints play an import role.

Failures due to out-of-bid situations may lead to the inability to provide the desired quality of service, e.g.: prolonged application execution times or an inability of applications to finish within a specified deadline. To overcome this uncertainty, one may come up with a few strategies to decrease the chance of failure or mitigate their effects.

To decrease the chance that out-of-bid situations occur, one could to choose to bid as high as possible. Given that, under the current model of Amazon spot instances, users pay at maximum the current spot price (not the actual bid), there would be no apparent disadvantages in bidding much higher than the spot price. However, there are incentives for adopting more aggressive bidding strategies, i.e. bidding close or even lower than the current spot price.

Firstly, Amazon offers on-demand instances at a fixed price, which are identically functional to spot instances and are not subject to terminations due to pricing issues. The value set by Amazon to these on-demand instances is likely to influence the maximum price a user is willing to bid. Thus, this value acts as an upper bound for bids of users that would rather lease a reliable on-demand instance in cases the spot price is equal or above the on-demand price. In fact, by analysing the history of spot prices of Amazon EC2, we have observed that, over the period of about 100 days from 05-Jul-2011 to 15-Oct-2011, spot prices have surpassed on-demand prices several times across most instances types and datacenters. For example, the spot price of one of the most economical instances (M1SMALL) in the US-EAST region, has reached this situation 11 times, for periods of up to 2 hours and 20 minutes, and price value of up to 17\% above the on-demand price.

Secondly, in a scenario where most users submit high bids, providers would likely increase the spot price to maximize profits. As previously postulated~\cite{mazzucco2011}, the Amazon EC2 spot market resembles a Vickrey auction style~\cite{ausubel2006lovely}, where users submit sealed bids, the provider gathers them and computes a clearing price. The pricing scheme thought to be used by Amazon, where all buyers pay the clearing price, is a generalization of the Vickrey model for multiple divisible goods, the standard uniform price auction, on which the provider assigns resources to users starting by the highest bidder, until all bids are satisfied or there are no more resources. The price paid by all users is the value of the lowest winning bid (sometimes, the highest non winning bid)~\cite{zhang2011}. Is has been observed that this scheme is a truthful auction, provided that the supply level is adjustable ex post (i.e. after the bids have been decided)~\cite{zhang2011}. It has also been observed that Amazon may be artificially intervening in the prices by setting a reserve price adn generating prices at random~\cite{ben2011deconstructing}. In any case, we argue that there is an incentive for users to submit fair bids, based on the true value they are willing to pay for the resource.

Thirdly, on a similar note, users may choose to postpone non-urgent tasks when prices are relatively high, hoping to obtain a lower price (the true value) later, a strategy that can be accomplished by placing a bid at the desired price and waiting for it to be fulfilled. Similarly, in the case of an out-of-bid situation, owners of a non-urgent task would prefer to wait for the request to be in-bid again, rather than obtaining a new resource under new lease terms (e.g. another VM type, or the same type at a higher bid).

Finally, as observed by Yi et al~\cite{yi2010reducing}, one can bid low to take advantage of the fact that the provider does not charge the partial hour that precedes an out-of-bid situation. Thus, delaying the termination of an instance, even when it is not needed, to the next hour boundary, one can expect a probability of failure before termination, potentially avoiding to pay for the last hour.

The choice of an exact bid value can be empirically derived from a number of factors, including observations of price history, the willingness of the user to run instances at less than a certain price or not run at all, and a minimum reliability level required. These factors, when reflected on the bid value, define how likely the system is able to meet time and cost constraints.

In any case, the adoption of more aggressive bidding strategies can result in more failures, and potentially undermine the cost savings, as a result of frequent loss of work. Therefore, resource provisioning policies aimed at running computational jobs on spot instances must be accompanied by fault mitigation techniques, especially tailored for the features of cloud computing spot instances. Notable features of spot instances may influence the way fault tolerance works in this scenario. Most notably, an hour-based billing granularity and non-payment of partial hours in the case of failures, guarantees payment of the actual progress of computation~\cite{yi2010reducing}. Additionally, given that providers, such as Amazon, freely provide a history of price variations, significantly more informed decisions can be made by observing the past behaviour.
\begin{figure*}[!ht]
\centering
\includegraphics[scale=0.9]{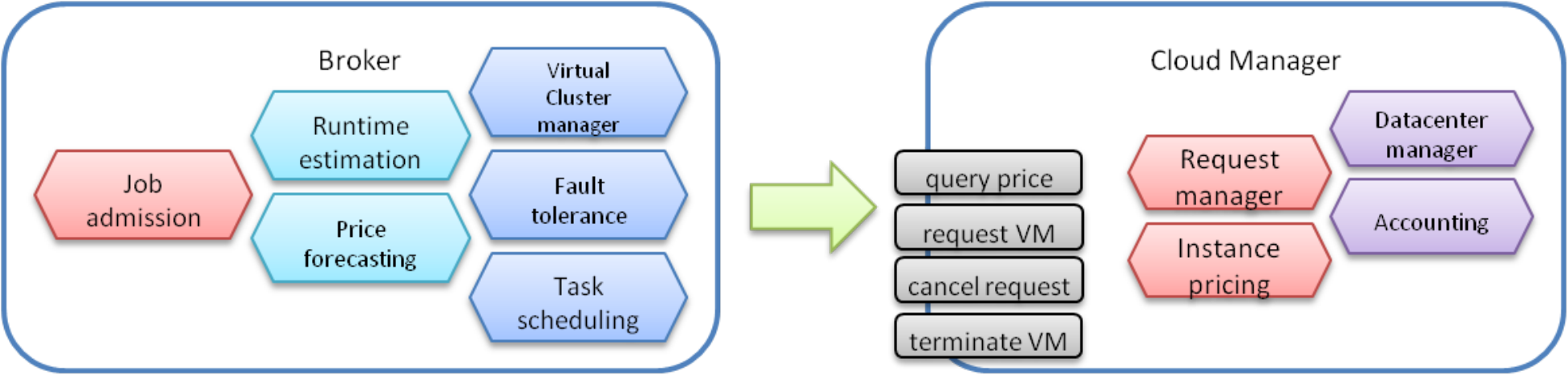}
\caption{Modeled architecture: Client (broker) and server (cloud) side. The ``Runtime estimation'' component was the focus of our previous work~\cite{voorsluys2011spot}. Here, we focus primarily on the ``fault tolerance'' component}
\label{fig:arch}
\end{figure*}
\subsection{Our contribution}
This paper proposes a resource provisioning strategy that addresses the problem of running computational jobs on intermittent VMs. Our main objective is to run applications in a fast and economical way, while tolerating sudden unavailability of virtual machines. We build up on our previous work~\cite{voorsluys2011spot}, where we demonstrated the viability of dynamically assembling virtual clusters exclusively composed of spot instances to run compute-intensive applications.

\textbf{Specifically, the contributions of this work are:}
\begin{itemize}
\item A multifaceted resource provisioning approach, that includes novel mechanisms for maximizing reliability, while minimizing costs in a spot instances-based computational platform;
\item An bidding mechanism that aids the decision-making process by estimating future spot prices and making informed bidding decisions;
\item An evaluation of two novel fault tolerance techniques, namely migration and job duplication, and their comparison to an existing checkpointing-based approach.
\end{itemize}

The rest of this paper is organized as follows: Section \ref{sec:related} describes related literature on existing approaches that use spot instances; Section \ref{sec:system_model} describes our existing resource provisioning policy and discusses the modifications necessary to add a reliability component to it; Section \ref{sec:mech} details our multifaceted approach and discusses each mechanism and the interaction between them; Section \ref{sec:eval} presents extensive simulation-based experimental results and their discussion; finally Section \ref{sec:conclusion} concludes the paper.

\section{Related Work}
\label{sec:related}

A few recently published works have touched the subject of leveraging variable pricing cloud resources in high-performance computing.
Andrzejak et al.~\cite{Andrzejak2010} have proposed a probabilistic decision model to help users decide how much to bid for a certain spot instance type in order to meet a certain monetary budget or a deadline. The model suggests bid values based on the probability of failures calculated using a mean of past prices from Amazon EC2. It can then estimate, with a given confidence, values for a budget and a deadline that can be achieved if the given bid is used.

Yi et al. \cite{yi2010reducing} proposed a method to reduce costs of computations and providing fault-tolerance when using EC2 spot instances. Based on the price history, they simulated how several checkpointing policies would perform when faced with out-of-bid situations. The proposed policies used two distinct techniques for deciding when to checkpoint a running program: at hour boundaries and at price rising edges. In the hour boundary scheme, checkpoints are taken periodically every hour, while in the rising edge scheme, checkpoints are taken when the spot price for a given instance type is increasing. The authors proposed combinations of the above mentioned schemes, including adaptive decisions, such as taking or skipping checkpointing at certain times. Their evaluation has shown that checkpointing schemes, in spite of the inherent overhead, can tolerate instance failures while reducing the price paid, as compared to normal on-demand instances. Similarly, we evaluate a checkpointing mechanism implemented according to this work, with the objective of comparing with other fault tolerance approaches.

\section{Resource Provisioning in a Spot Instances-based Computational Platform}
\label{sec:system_model}

In our previous work, we have proposed a resource provisioning and job allocation architecture and an associated policy. Our solution has been tailored for an organization that aims at assembling a computational platform solely based on spot instances and use it to accomplish a stream of deadline-constrained computational jobs. In that work, we also evaluated several runtime estimation mechanisms and their effect on cost and utilization of the platform, as well as deadline violations of jobs. In this section, we summarize how our solution works; a detailed description and analysis can be found in~\cite{voorsluys2011spot}.

A Broker component is responsible for receiving computational job requests from users, provisioning a suitable VM pool by interacting with the provider, and applying a job scheduling policy to ensure jobs finish within their deadlines, while minimizing the cost. A diagram depicting the components of the modeled architecture is shown in Figure \ref{fig:arch}.

We have modeled a cloud computing provider according to how Amazon EC2 currently works in practice. The provider manages a computational cloud, formed by one or more datacenters, which offer virtual machines of predefined types in a spot market. The provisioning of an instance is subject to the following characteristics: clients submit requests for a single instance, specifying a type, and up to how much they are willing to pay per instance/hour (bid). Optionally, a particular datacenter can be specified; if left blank, the provider allocates the instance to the most economical datacenter choice. The system provides instances whenever the bid is greater than the current price; on the other hand, it terminates instances without any notice when a client's bid is less than or equal to the current price. The system does not charge the last partial hour when it stops an instance, but it charges the last partial hour when the termination is initiated by the client (the price of a partial hour is considered the same as a full hour). The price of each instance/hour is the spot price at the beginning of the hour.

Jobs are assumed to be moldable, in the sense that they can run on any number of CPU cores, but limited to a single virtual machine. To determine the run time of a job in a particular number of CPU cores, we use Downey's analytical model for job speedup \cite{downey1997}. To generate values for $A$ (average parallelism) and $\sigma$ (coefficient of variance of parallelism), we have used the model of Cirne \& Berman \cite{cirne2001}. The moldability of a job defines it's preferred instance type, i.e. the type on which the job will take advantage of the most number of cores for a time greater than 1 hour. As a result, longer jobs that offer more parallelism will prefer instances with more cores.

The activities of our proposed algorithm are summarized in the steps described below.

\begin{itemize}
\item When any job is submitted, it is inserted into a list of unscheduled jobs;
\item At regular intervals ($T$), the algorithm uses a runtime estimation method to predict the approximate runtime of the job on each available instance type;
\item The broker then attempts to allocate the job to an idle VM with enough time before a whole hour finishes;
\item If unsuccessful, it attempts to allocate the job to a VM that is currently running jobs but is expected to become idle soon. Runtime estimates of all jobs running on the VM, in addition to the incoming job, are required at this step;
\item If the job still cannot be allocated, the algorithm will decide whether it is advantageous to extend a current lease, to start a new VM lease, or to postpone the allocation decision according to the job's urgency factor and pricing conditions.
\end{itemize}

The urgency factor $U$ of a job $j$ is the maximum estimated time the job can wait for a resource to be provisioned so that the chance of meeting the deadline is increased. It is computed as per Equation \ref{eq_alpha}, where $D_{j}$ is the job's deadline, $T$ is the current time, so that $D_{j} - T$ corresponds to the time until the job's deadline; $\alpha$ is the urgency modifier; $e_{j}$ is the estimated runtime of $j$ on it's preferred instance type; and $B$ is the expected time the provider takes to provision a new VM (fixed at 5 minutes).

\begin{equation}
\label{eq_alpha}
U_{j} = max(0, D_{j} - T - (\alpha * e_{j} + B))
\end{equation}

The greater the value of the $\alpha$ modifier, the more conservative the algorithm becomes, i.e. with higher values of $\alpha$, $U$ approximates $0$. A value equal to $0$ indicates that a resource must be provisioned immediately to complete the job within the deadline. Alternatively, lower values of $\alpha$ cause the algorithm to postpone more provisioning actions in order to maximise the chances of finding lower prices or reusing other jobs' instances.

\section{Mechanisms to Achieve Fault Tolerance}
\label{sec:mech}

In this work, we explore a multifaceted approach, which relies on two interrelated modalities that define how reliably the policy ensures that computational jobs finish before their deadlines. The first mechanism aims at choosing appropriate bid values based on estimation of price variations and on the job's urgency factor $U$, which influences the choice of when to provision a resource for a given job and how much to bid. The second mechanism adds extra levels of fault tolerance through checkpointing and migration of virtual machines, as well as job duplication.

These mechanisms aim at mitigating spot instance unavailability due to out-of-bid situations only, i.e. failures due to price variations. Other types of instance failures, for instance, due to hardware faults or network interruptions are not considered. In other words, we assume that, if no out-of-bid situation takes place during an instance lifetime, its availability is 100\%.

\subsection{Bidding strategies: estimating cost and jobs' urgency}

The first mechanism comprises bidding strategies and the calculation of the value of $U$. These are based on estimated price variations and job runtimes. More specifically, this mechanism aims to aid the process in two ways: (1) allow the broker to make informed decisions on how much to bid, a choice that directly influences the risk of failure and monetary spending; and (2) combine price information and a job's urgency factor, to decide the best point in time to start a new machine for a job, thus seeking to cover the period that will yield the minimum cost. The rationale behind combining these two pieces of information is to avoid hasty decision that may increase costs, i.e. to avoid commissioning new resources too early, at times when non urgent jobs can be postponed, or too late, when jobs will most likely miss their deadlines.

In our previous work~\cite{voorsluys2011spot}, we have compared several runtime estimation policies and their impact on cost, deadline violations, and system utilization. A simple mechanism that computes the average runtime of two preceding jobs of the same user has performed consistently well. Therefore, in this work, we exclusively employ that technique.

We have evaluated 5 bidding strategies, which are listed on table \ref{tab:bidding}. Two of the strategies use historical information to compute the bid. In all cases, a window of one week worth of price history, individual to each instance type/OS/datacenter combination, is fed to the bidding strategy. The output of each strategy is the maximum price, in US dollars per hour, to be paid for one particular instance. The minimum bid granularity $G$ is 0.001.

\begin{table}
\centering
\caption{Evaluated bidding strategies}
\label{tab:bidding}
\scriptsize
\begin{tabular}{|l| l|}
		\hline
		Bidding strategy & Bid value definition \\
		\hline
		Minimum & The minimum value observed in the price history + $G$ \\
		Mean & The mean of all values in the price history \\
		On-demand & The listed on demand price \\
		High & A value much greater than any price observed (defined as 100)\\
		Current & The current spot price + $G$ \\
		\hline
\end{tabular}
\end{table}

In all cases that can yield values lower than the current price, the broker uses the value of $U$ to override the bid value, if necessary. Specifically, it applies the steps of Algorithm \ref{algo:bidding}.

\scriptsize
\begin{algorithm}
  \BlankLine

  $b \leftarrow$ compute bid\;
  $U \leftarrow$ compute urgency factor\;
  $P \leftarrow$ query provider for current price\;
  \eIf{$U = 0$}{
   \If{$b <= P$}{
    $b = P + G$\;
  }}{
    schedule a bid check at $T + U$\;
  }
  \BlankLine
\caption{Bid check algorithm, which overrides the bid value or schedules a new check in the future}
\label{algo:bidding}
\end{algorithm}
\normalsize

\subsection{Hourly Checkpointing}

Checkpointing consists of saving the state of a VM, application, or process, during execution and restoring the saved state after a failure to reduce the amount of lost work~\cite{Kintala1995}. In the context of virtual machines, the action of encapsulating execution and user customization state is a commonplace feature in most virtual machine monitors (VMM)~\cite{Lee2011}. Saving a VM state consists of serializing its entire memory contents to a persistent storage, thus including all applications and process running~\cite{Kozuch2002}. In our work, we assume that checkpointing a running application is the same as saving the state of an entire VM. The advantage of relying on VMM-supported checkpointing is that applications do not need to be modified to enable checkpointing-based fault tolerance. However, it is necessary that cloud computing providers explicitly support such operation.

The technique considered in this work is a hourly-based VM checkpointing, where states are saved at hour-boundaries. This technique has been previously identified by Yi et al.~\cite{yi2010reducing} as the simplest and most intuitive, yet effective, form of dealing with the cost/reliability trade-off when running applications on spot instances. More specifically, taking a checkpoint on an hourly basis guarantees that only useful computational time is paid, given that spot instances are billed at an hour granularity and partial hours, in the case of failures, are not charged.

In this method, it is assumed that a checkpointed VM will only resume when the original spot request, which has a fixed bid and machine type, is in-bid again. No attempt is made to provision a new VM by submitting higher bids for the same machine type, or to bid for other types. This contrasts with our next solution, which considers relocating the saved state to a new space in order to hasten job completion.

\subsection{Migration of persistent VM state}

We propose a migration-based fault tolerance mechanism on which the state of a VM is frequently saved on a global filesystem and upon an out-of-bid situation the state is relocated. The migration technique is very similar to checkpointing, as it comprises of taking a snapshot of the VM and using it to restore the computation upon a failure. But instead of waiting for the original request to be in-bid again, the algorithm aims to lease a new instance under new terms, and then restore the saved VM state into the new instance.

The definition of new lease terms is subject to the following decision (whichever is estimated to be cheaper to accomplish the remaining duration of the job): (1) leasing an instance of the same type for a higher price in the same datacenter; (2) leasing an instance of a different type on the same datacenter; (3) or relocating the workload to another datacenter where a suitable VM may be leased for a cheaper price. The overhead of restoring a failed VM in a distinct datacenter is assumed to be higher than when the same datacenter is chosen. This overhead is taken into account by the algorithm when making a relocation decision.

All computation in the VM is paused while the snapshot is being taken. The overhead of saving an instance state (the same as taking a checkpoint) is defined as the time to serialize a VM's memory snapshot into a file in a global filesystem. This value is different for each instance type, according to their maximum memory size. The exact values are computed as in the work of Sotomayor et al.~\cite{Sotomayor2009}, which provides a comprehensive model to predict the time to suspend and resume VMs. The times to suspend (i.e. save the state) and to resume (i.e. restore from the latest saved state) a spot instance with $m$ MB of memory, are defined as per equations \ref{eq_ts} and \ref{eq_tr} respectively~\cite{Sotomayor2009}.

\begin{equation}
\label{eq_ts}
t_{s} = m/s
\end{equation}
\begin{equation}
\label{eq_tr}
t_{r} = m/r
\end{equation}

Values for $s$ and $r$ (rates, in MB/s, to write/read $m$ MB of memory to/from a global filesystem) are also taken from~\cite{Sotomayor2009}, who obtained them from numerous experiments on a realistic testbed. Therefore, $s$ is 63.67 MB/s, and $r$ is 81.27 MB/s (to restore a state in the same datacenter). We assume half the rate (40.64 MB/s) when moving/restoring a VM state into/from a distinct datacenter.

\begin{figure*}[ht]
\centering
\subfigure[Monetary cost]{\includegraphics[scale=0.3]{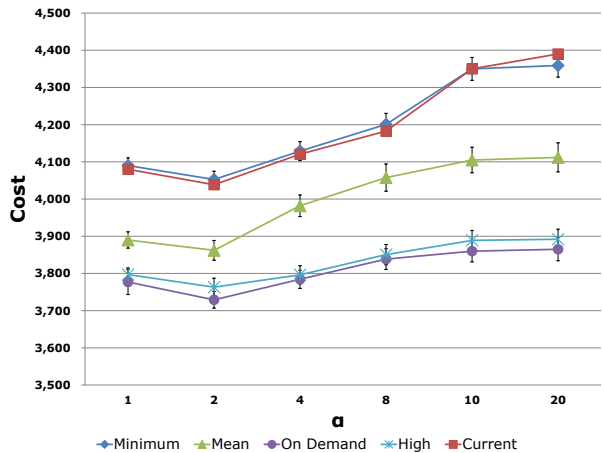}\label{fig:urg_vs_cost}}
\subfigure[Deadline violations]{\includegraphics[scale=0.3]{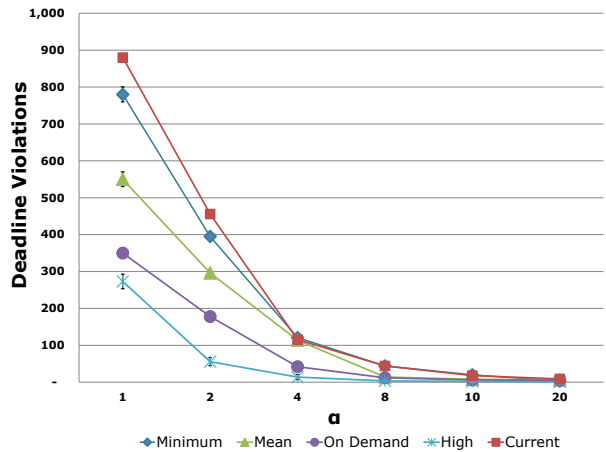}\label{fig:urg_vs_ddl}}
\caption{Effect of aggressive and conservative urgency estimation modifier ($\alpha$) under various bidding strategies}
\label{fig:none}
\end{figure*}

\subsection{Duplication of long jobs}

We also propose a fault tolerance mechanism that does not require any application- or provider-assisted technique, as it is the case of VM-based checkpointing and migration. With task duplication, we aim to evaluate a simpler method for rapid deployment of applications on spot instances using currently available cloud computing feature.

Similar to replication and migration, duplication of work aims to increase the chance of success in meeting deadlines when running longer jobs (greater than one hour) over a period of frequent price changes. Therefore, a duplication-based technique was implemented and evaluated.

This technique also relies on estimates of jobs runtimes. It creates one replica of each job that is expected to run for more than 1 hour. The replica is submitted to the same scheduling policy as the original job. The algorithm applies the same rules as it does to a regular job, but avoids choosing the the datacenter/type combination where the original job will run. Choosing a different combination for a replica is an obvious choice, since two jobs running on the same datacenter, using the same instance type, will certainly fail at the same time when the price increases.

\section{Performance Evaluation}
\label{sec:eval}

In this section, we evaluate the proposed fault-aware resource allocation policy and the effect of its mechanisms, using trace-driven discrete event simulations. We quantify the performance of our policy based on three metrics, two absolute (monetary cost and deadline violations) and one relative (dollar per useful computation). We especially observe the interaction between these metrics, given that there is a known trade-off between them, i.e. assuring less violations usually means provisioning more resources, hence higher costs.

\subsection{Experimental design}

We have designed our experiments to study the influence of the following factors and their levels: (1) bidding strategy; (2) the value of the urgency factor modifier $\alpha$; and (3) choice of fault tolerance mechanism. The factors and their levels are listed on Table \ref{tab:factors}.

\begin{table}[t]
\centering
\caption{Factors and their levels}
\label{tab:factors}
\scriptsize
\begin{tabular}{|l|l|}
		\hline
		Factor & Possible values \\
		\hline
		Bidding Strategy & Minimum, Mean, On-demand, High, Current \\
		$\alpha$ & 1, 2, 4, 8, 10, 20 \\
		Fault tolerance mechanisms & None, Migration, Checkpointing, Job duplication \\
		\hline
\end{tabular}
\end{table}

Not all combinations of factors have been simulated; for example, these was little sense in combining the High bidding strategy with a fault tolerance mechanism, given that the bidding fashion itself completely avoid failures. In total, 5952 experiments were executed. All values presented correspond to an average of 31 simulation runs. When available, error bars correspond to a 95\% confidence interval. The simulator was implemented using the CloudSim framework~\cite{buyya2009modeling}.

\subsubsection*{Cloud characteristics} We modeled the cloud provider after the features of Amazon EC2's US-EAST geographic region, which contains 4 datacenters. Instance types were modeled directly after the characteristics of available standard and high-CPU types The types available to be used are M1.SMALL (1 ECU), M1.LARGE (5 ECUs), M1.XLARGE (8 ECUs), C1.MEDIUM (5 ECUs), C1.XLARGE (20 ECUs). One ECU (EC2 Compute Unit) is defined as equivalent to the power of a 1.0-1.2 GHz 2007 AMD Opteron or 2007 Intel Xeon processor. A period of 100 days worth of pricing history traces has been collected comprising dates between July 5th, 2011 and October, 15th, 2011. These dates correspond to the available traces since Amazon EC2 has started offering distinct prices per individual datacenter, rather than per geographic region.

\subsubsection*{Workload} The chosen job stream was obtained from the LHC Grid at CERN~\cite{feitelson71parallel}, and is composed of grid-like embarrassingly parallel tasks. A total of 100,000 jobs are submitted over a period of seven days of simulation time, starting from a randomly generated time within the available price history. This workload is suitable to our experiments to due to its bursty nature and for being composed of highly variable job lengths. These features require a highly dynamic computation platform that must serve variable loads while maintaining cost efficiency. The moldability parameters $A$ and $\sigma$ of each job are assumed to be known by the broker. 

Originally, this workload trace did not contain information about user-supplied job runtime estimates and deadlines. User runtime estimates were generated according to the model of Tzafrir et al.~\cite{Tsafrir05modelinguser}. A job's maximum allowed runtime corresponds to the runtime estimate multiplied by a random multiplier, uniformly generated between 1.5 and 4. Consequently, the job's deadline corresponds to its submission time plus its maximum allowed runtime.

\subsection{Effects of bidding strategies and urgency factor}

\begin{figure*}[ht]
\centering
\subfigure[Migration]{\includegraphics[scale=0.21]{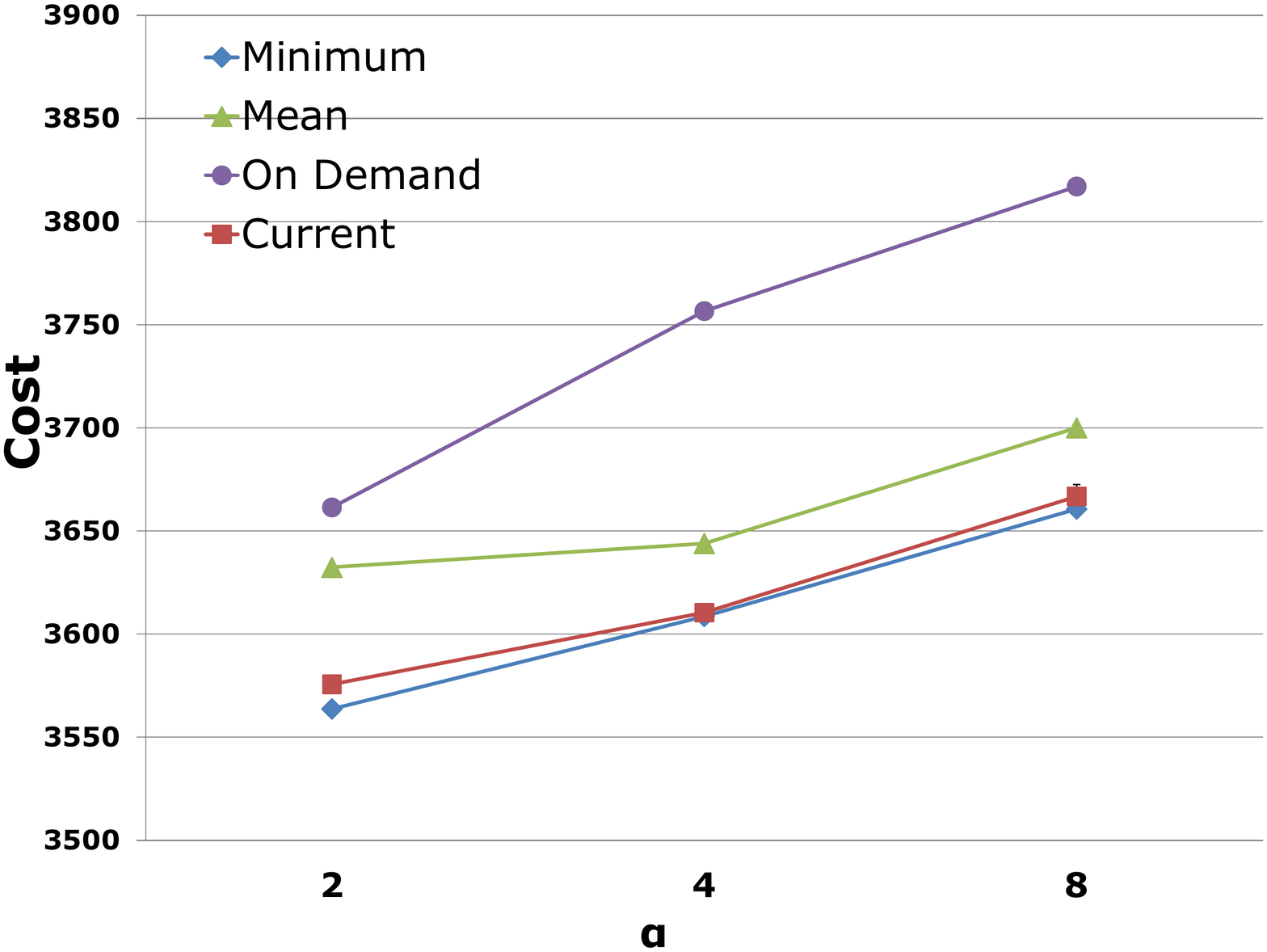}\label{fig:mig1}}
\subfigure[Checkpointing]{\includegraphics[scale=0.21]{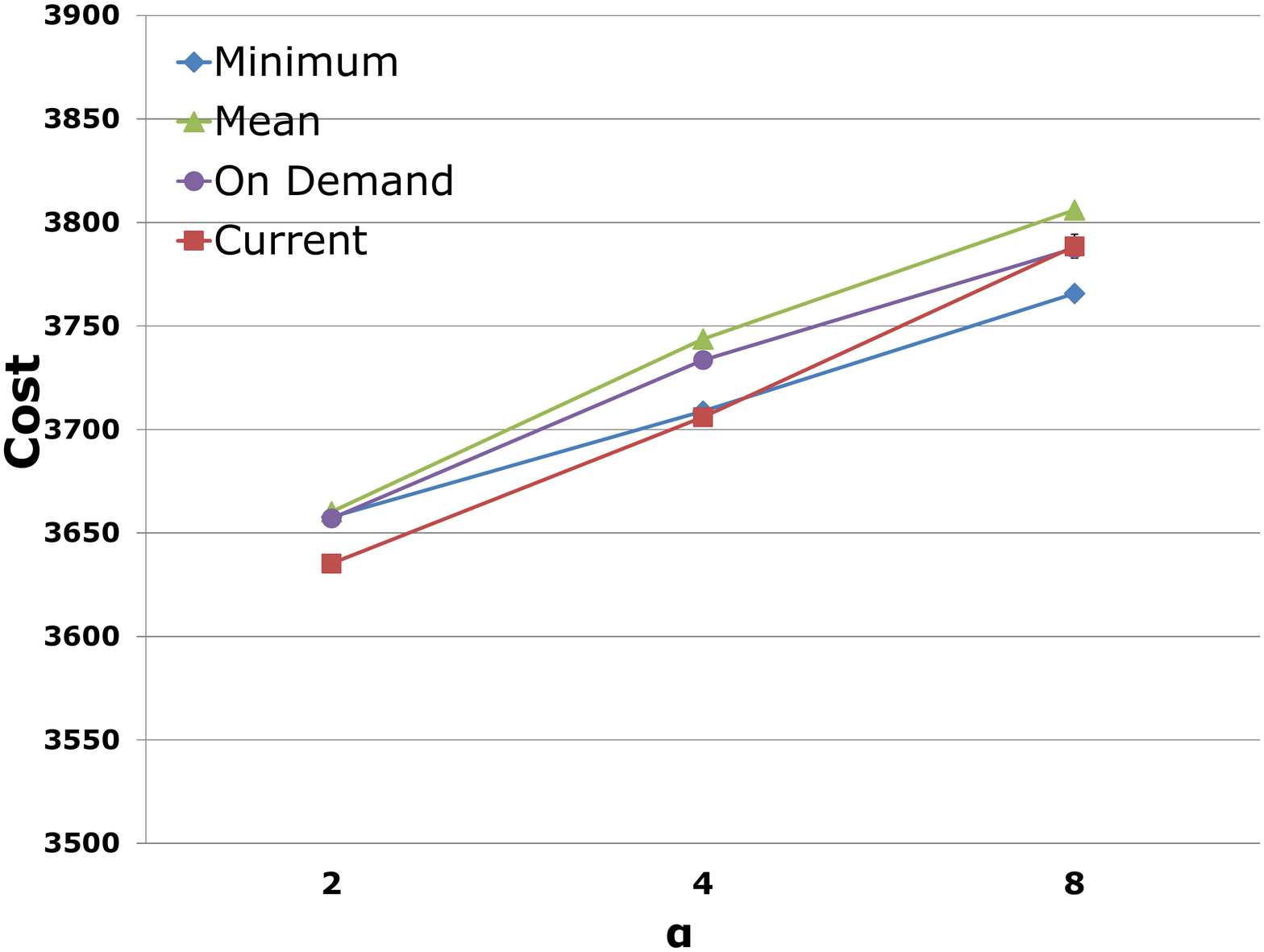}\label{fig:chk1}}
\subfigure[Job duplication]{\includegraphics[scale=0.21]{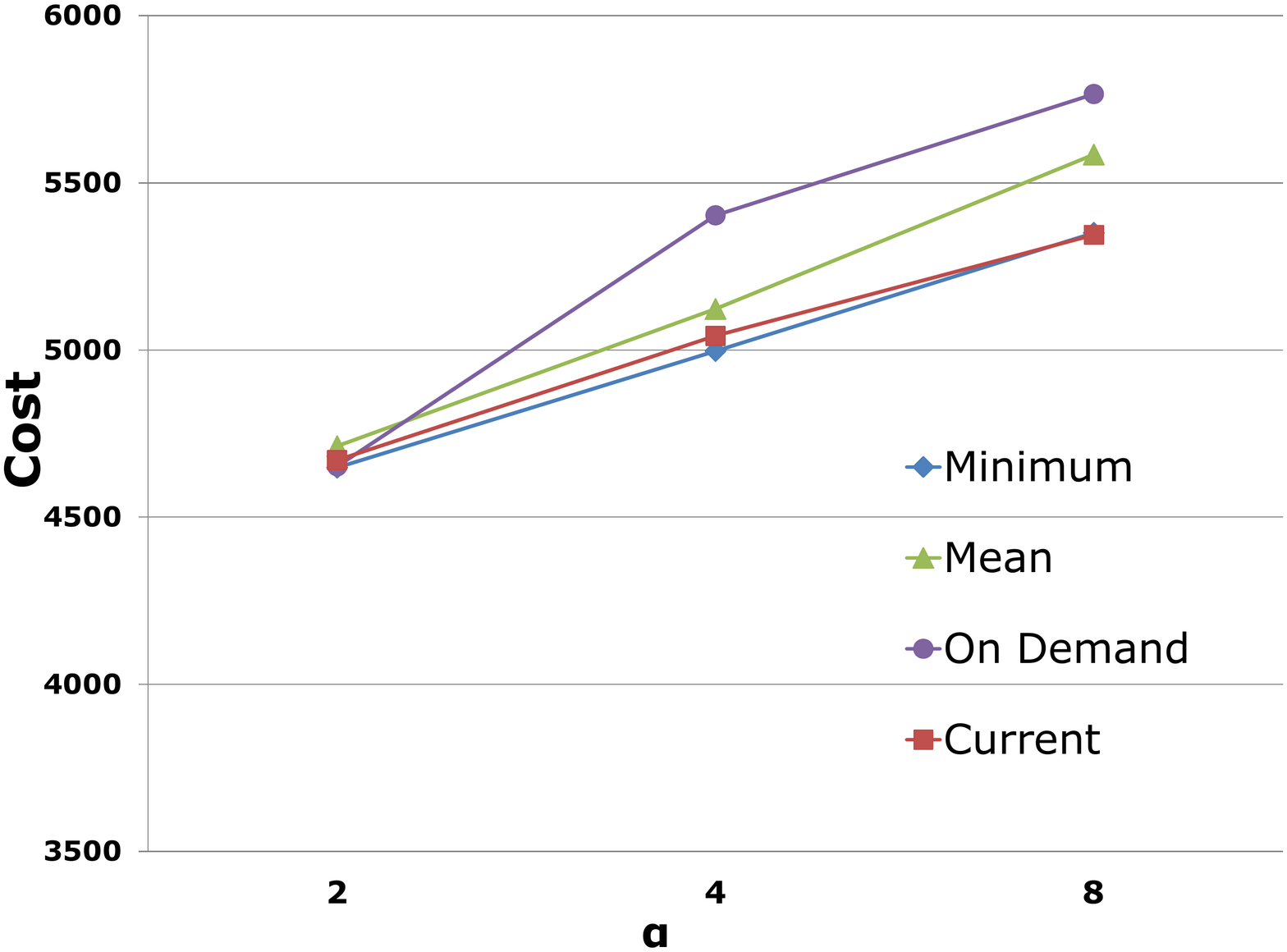}\label{fig:repl1}}
\caption{Performance of migration, checkpointing and job duplication on monetary cost}
\label{fig:ft}
\end{figure*}

In order to understand how our bidding strategies work, independently of fault tolerance mechanisms, we have evaluated their effectiveness in a scenario where a failed job must be restarted from the beginning after a failure. In this experiment, we aimed at quantifying each strategy's performance when paired with different values of $\alpha$.

Figure \ref{fig:none} shows the effect of most aggressive ($\alpha = 1$) to most conservative ($\alpha = 20$) urgency estimations under various bidding strategies. In these circumstances, bidding strategies that produce higher bids tend to perform better, both in terms of cost and deadline violations. In particular, we have observed that the On-demand strategy avoids failures due to minor price increases, as well as avoids incurring the cost of high prices above the on-demand price. This fact can be noticed in the performance comparison between On-demand and High, which incurs extra cost due its very high bid. As expected, strategies that aim at bidding low values experience the most failures, hence more loss of work, and consequently higher costs.

The value of $\alpha$ significantly influences both cost and deadline violations, consistently over all bidding strategies. Figure \ref{fig:urg_vs_cost}, indicates an optimal value of 2, which yields the lower costs, although for most bidding strategies, the difference between 1 and 2 is not statistically significant. Regarding the deadline metric, 1 and 2 lead to many more deadline violations. This is due to the fact that lower values of $\alpha$ cause the algorithm to postpone more decisions, which in turn often leads to the inability of provisioning resources ``at the last minute''. Conservative values, on the other hand, lead to virtually no violations, but significantly higher costs.

\subsection{Migration, checkpointing, and job duplication}

Our results also demonstrate the positive effects of the studied fault tolerance mechanisms when paired with bidding strategies and urgency factor estimation. Figure \ref{fig:ft} shows a comparison of migration, checkpoint and job duplication on the cost metric. We only show values of $\alpha$ of 2, 4 and 8, which yield the best costs in all cases. An interesting fact is that migration performs better when paired with bidding strategies that choose lower bid values, such as Minimum and Current, while checkpointing benefits from higher bid values, such as Mean and On-demand. This behaviour is coherent with the features of each mechanism. Migration tends to have more choices after an out-of-bid situation given its ability to choose other types of instances from multiple datacenters. Checkpointing, on the other hand, is bound to a persistent request, and will benefit from a higher chance of being in-bid most of the time.

Job duplication performs poorly in all cases, yielding much higher costs when compared to the case when no fault tolerance exists. It's merit however, lies on its simplicity and the capability of replicating jobs across multiple datacenter. Therefore, it can be useful in cases where an extra level of redundancy is required.

\begin{figure}[!b]
\centering
\includegraphics[scale=0.3]{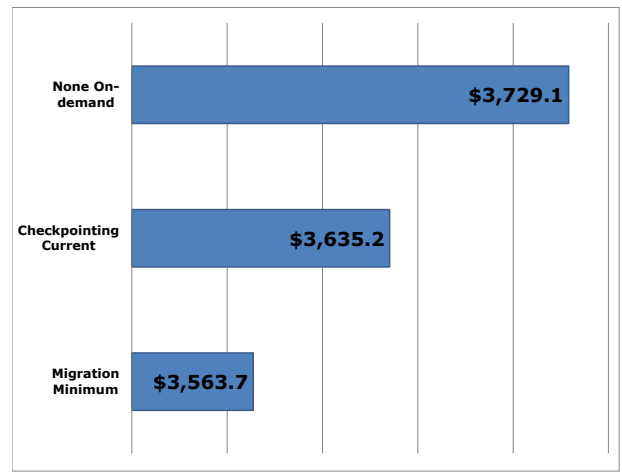}
\caption{Most economical combinations of bidding strategy and fault tolerance mechanism}
\label{fig:comp}
\end{figure}

Figure \ref{fig:comp} presents a summary of best combinations of strategies discovered in our simulations. Overall, the migration technique, along with the Minimum bidding strategy and $\alpha = 2$ produced the lower cost. However, $\alpha = 8$ produced the least number of deadline violations (30 out of 100,000 jobs). These results confirm that the trade-off between cost and deadline violations applies in this case.

In summary, our results demonstrate that the interaction of factors can influence the exact choice of bidding strategy, $\alpha$, and fault tolerance mechanism. It is expected that, in absolute terms, more conservative urgency factors will lead to less deadline violations and a greater cost. To help gauge a more precise metric, we define dollars per useful computation as the ratio between the total cost and the number of jobs that finished within their deadlines. Table~\ref{tab:rank} ranks the 10 best factor combinations according this metric. The combinations that employ migration rank consistently superior, which makes these combinations good candidates for environments where a strict meeting of deadlines is expected.

\begin{table}
\centering
\caption{Analysis of the dollars per useful computation metric}
\label{tab:rank}
\scriptsize
\begin{tabular}{|p{0.5cm}|p{1.5cm}|p{1.3cm}|l|p{1.3cm}|p{1.5cm}|}
\hline
Rank & Fault tolerance & Bidding strategy & $\alpha$ & Dollars per useful computation & Worsening related to best (\%)\\
\hline
1&Migration&Minimum&2&0.03578&0\\
2&Migration&Current&2&0.03588&0.29\\
3&Migration&Minimum&4&0.03613&0.978\\
4&Migration&Current&4&0.03614&1.006\\
5&Migration&Mean&2&0.03641&1.736\\
6&Checkpointing&Current&2&0.03647&1.881\\
7&Migration&Mean&4&0.03648&1.932\\
8&Migration&Minimum&8&0.03661&2.279\\
9&Checkpointing&On-demand&2&0.03663&2.330\\
10&Checkpointing&Mean&2&0.03666&2.412\\
...& & & & &\\
18&None&On-demand&2&0.03736&4.224\\
\hline
\end{tabular}
\end{table}

\section{Conclusions and Future Work}
\label{sec:conclusion}

In this work, he have proposed a multifaceted resource provisioning policy that reliably manages a pool of intermittent spot instances. Our policy contains multiple mechanisms, including 5 bidding strategies, an adjustable urgency factor estimator, and 3 fault-tolerance approaches. 

We have performed extensive simulations under realistic conditions that reflect the behaviour of Amazon EC2, via a history of its prices.
Our results demonstrate that both costs savings and stricter adherence to deadlines can be achieved when properly combining and tuning the policy mechanisms. Especially, the fault tolerance mechanism that employs migration of VM state providers superior results in virtually all metrics.

Currently, the cloud computing spot market is still in its infancy. Therefore, many challenges have not been encountered, given the short history and relatively low variability of Amazon EC2 prices. In this sense, we plan to further improve our policy by devising bidding strategies that will perform well in environments with highly variable price levels and more frequent changes. We expect fault tolerance to be even more crucial in such scenarios. We also plan to lean towards provider-centric research, by studying the challenges involved in setting spot prices under various demand patterns.

\subsubsection*{Acknowledgements}

This work is partially supported by grants from the Australian Research Council (ARC), and the Australian Department of Innovation, Industry, Science and Research (DIISR). We also would like to thank ej-technologies GmbH for providing a license of jProfiler for use in the CloudSim project.

\bibliographystyle{IEEETran}

\bibliography{spotsim}

\end{document}